\documentclass{article}
\usepackage{emulateapj,apjfonts,rotating,epsf}

\lefthead{Ralf\ S.\ Klessen}
\righthead{Time-Varying Accretion Rates}

\begin{document}

\title{The Formation of Stellar Clusters: Time Varying Protostellar Accretion Rates}
 \author{Ralf S.\ Klessen}
\affil{UCO/Lick Observatory, University of California at
 Santa Cruz, Santa
 Cruz, CA 95064, U.S.A. \\ Otto Hahn Fellow, sponsored by  Max-Planck-Institut f{\"u}r
 Astronomie, K{\"o}nigstuhl 17, 69117 Heidelberg, Germany}

\begin{abstract}
  Identifying the processes that determine strength, duration and
  variability of protostellar mass growth is a fundamental ingredient
  of any theory of star formation. I discuss protostellar mass
  accretion rates $\dot{M}$ from numerical models which follow
  molecular cloud evolution from turbulent fragmentation towards the
  formation of stellar clusters. In a dense cluster environment,
  $\dot{M}$ is strongly time varying and influenced by the mutual
  interaction of protostellar cores and their competition for
  accretion from the common cluster gas reservoir. Even for protostars
  with similar final mass, the accretion histories may differ
  dramatically. High-mass stars build up in the central parts of
  dense, cluster-forming cloud regions. They begin to form in the
  early phases of cluster evolution, and continue to grow at a high
  rate until the available gas is exhausted or expelled by feedback.
  Lower-mass stars tend to form at later phases, and $\dot{M}$
  declines rapidly after a short initial phase of strong growth.  I
  present a simple fit formula for the time evolution of the average
  $\dot{M}$ for protostars of different masses in a dense cluster
  environment.
\end{abstract}


\section{Introduction}
Star formation is a complex process. In the standard theory of star
formation, stars build up from the ``inside-out'' collapse of singular
isothermal spheres, which are generally assumed to result from the
quasistatic contraction of magnetically supported cloud cores due to
ambipolar diffusion (Shu 1977, Shu, Adams, \& Lizano 1987). This
picture is able to reproduce many observed features of protostars,
however, has always been challenged by more dynamical points of view
(see e.g.\ Whitworth et al.\ 1996 for a critical discussion). An
alternative approach takes interstellar turbulence into account, where
supersonic random motions in molecular clouds lead to transient
shock-generated density fluctuations. Some may exceed the threshold of
gravitational stability and collapse to form stars (e.g.\ Elmegreen
1993, Padoan 1995, Padoan \& Nordlund 1999). The location, timescale
and efficiency of star formation hereby intimately depend on the
stochastic properties of the underlying turbulent velocity field
(Klessen, Heitsch, \& Mac Low 2000, Heitsch, Mac Low, \& Klessen 2001,
hereafter KHM and HMK, respectively, also Klessen 2000). The majority of stars form in a
clustered environment (see the review by Clarke, Bonnell, \&
Hillenbrand 2000, and Elmegreen et al.\ 2000). For massive clusters, 
the mutual interaction of collapsing protostellar cores in the dense central
region and their competition for accretion from a common gas reservoir
are important processes which determine stellar mass growth
(Bonnell et al.\ 1997, Klessen \& Burkert 2000, 2001, hereafter, KB1
and KB2, respectively). It is this star forming environment which is
considered in the present paper.

A gravitationally unstable, collapsing gas clump follows an
observationally well determined sequence.  Prior to the formation of a
hydrostatic nucleus, an observed pre-stellar condensation exhibits a
density structure which has a flat inner part, then decreases outward
roughly as $\rho \propto r^{-2}$, and is truncated at some finite
radius (e.g.\ Bacmann et al.\ 2000). Once the central YSO builds up,
the class 0 phase is reached and the density follows $\rho \propto
r^{-2}$ down to the observational resolution limit. As larger and
larger portions of the infalling envelope get accreted the protostar
is identified as class I object, and when accretion fades away it
enters the T Tauri phase (e.g.\ Andr{\'e}, Ward-Thompson, \& Barsony
2000).  IR and sub-mm observations suggest that the accretion rate
$\dot{M}$ varies strongly and declines with time. Accretion is largest
in the class 0 phase and drops significantly in the subsequent
evolution (e.g.\ Andr{\'e} \& Montmerle 1994, Bontemps et al.\ 1996,
Hendriksen, Andr{\'e}, \& Bontemps 1997, hereafter HAB). The estimated
lifetimes are a few $10^4\,$years for the class 0 and a few
$10^5\,$years for the class I phase.

These observational findings favor the dynamical picture of star
formation (e.g.\ Larson 1969, Penston 1969, Hunter 1977, HAB, Basu
1997), but could in principle be reconciled with an "inside-out"
collapse model when adopting initial conditions differnt from the
classical quasi-static singular isothermal sphere. All analytical
studies and most numerical work of protostellar core collapse (e.g.\ 
Foster \& Chevalier 1993, Tomisaka 1996, Ogino, Tomisaka, \& Nakamura
1999, Wuchterl \& Tscharnuter 2001, herafter WT) concentrate on
isolated objects. It is thus the aim of the present study to
illustrate the effect of a dense cluster environment on protostellar
mass accretion rates.

\section{Numerical Models of Clustered Star Formation}
In the approach adopted here, molecular clouds form clusters of stars
in regions where interstellar turbulence becomes too weak to supply
sufficient support against gravitational collapse, either because the
region is compressed by a large-scale shock (KHM, HMK), or because
interstellar turbulence is not replenished and decays on short
timescales (Mac~Low et al.\ 1998, Stone, Ostriker, \& Gammie 1998, Mac
Low 1999). The latter scenario is investigated by Klessen, Burkert, \&
Bate (1998), KB1, and KB2, using smoothed particle hydrodynamics (SPH)
simulations of the evolution and fragmentation of molecular cloud
regions where decaying turbulence is assumed to have left behind
random Gaussian density fluctuations. The complete lack of turbulent
support leads to rapid star formation in a strongly clustered
mode. The interplay between self-gravity and isothermal gas pressure
considered in the models reproduces the basic features of young star
clusters remarkably well, e.g.\ the mass distribution of protostars is
well fit by a log-normal with width and peak comparable to
observational estimates of the IMF (KB1, KB2).  The current analysis
is based on the high-resolution simulation $\cal I$ discussed in KB1.

The considered molecular cloud region has a mean density of $n({\rm
H}_2) = 10^5\,$cm$^{-3}$ and a temperature of 10$\,$K. Its initial
Gaussian density field follows a power spectrum $P(k)\propto k^{-2}$
in a volume of $\left(0.32\,{\rm pc}\right)^3$ containing roughly
$200\,$M$_{\odot}$. As the system contracts gravitationally, a cluster
of 56 protostellar cores builds up. As no feedback is included, the
common cluster gas reservoir from which these cores feed is exhausted
after roughly $4\times 10^5\,$years, corresponding to $\sim 3$ global
free-fall timescales.
  
Once the nucleus of a collapsing gas clump in the model exceeds a
density $n({\rm H}_2) \approx 10^{9}\,$cm$^{-3}$, it is identified as
protostellar core and replaced by a `sink' particle, which has the
ability to accrete matter while keeping track of mass, linear and
angular momentum (Bate, Bonnell, \& Price 1994).  Matter that accretes
through the boundary of the `sink' volume would continue to fall and
reach the central hydrostatic YSO within less than $\sim 1000\,$years
(WT).  The diameter of the accretion volume is $320\,$AU, roughly
equivalent to the Jeans scale at the threshold density. For the
stellar mass range considered here feedback effects are not strong
enough to halt or delay accretion onto the stellar photosphere
(Wuchterl \& Klessen 2001, hereafter WK).  Thus, the core accretion
rates derived here are good estimates for the actual stellar accretion
rates.  Deviations are expected only for very high-mass stars, which
will at some stage begin to emit UV photons, or for protostellar cores
forming binary stars, where the infalling mass must be distributed
between two stars, or if very high-angular momentum material is
accreted, where a certain mass fraction may end up in a circumbinary
disk and not accrete onto a star at all.

\begin{figure*}[p]
\unitlength1cm
\begin{picture}(18.0,23.9)
\put( 22.0, -4.3){\begin{rotate}{90} \epsfxsize=20.5cm \epsfbox{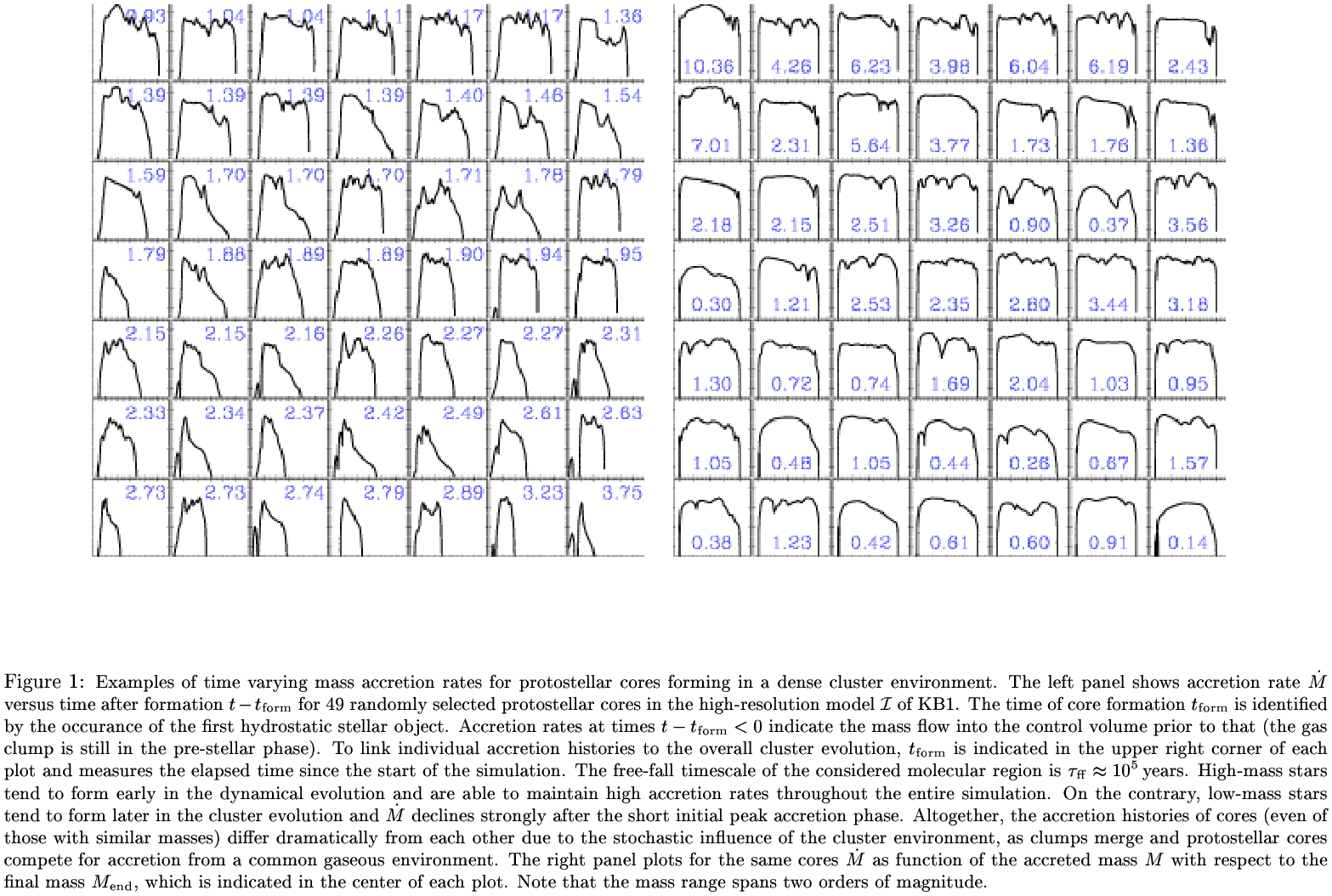} \end{rotate}}
\end{picture}
\end{figure*}
\begin{figure*}[th]
\unitlength1cm
\begin{picture}(18.0,12.7)
\put(-3.5, -7.4){\epsfxsize=20.5cm \epsfbox{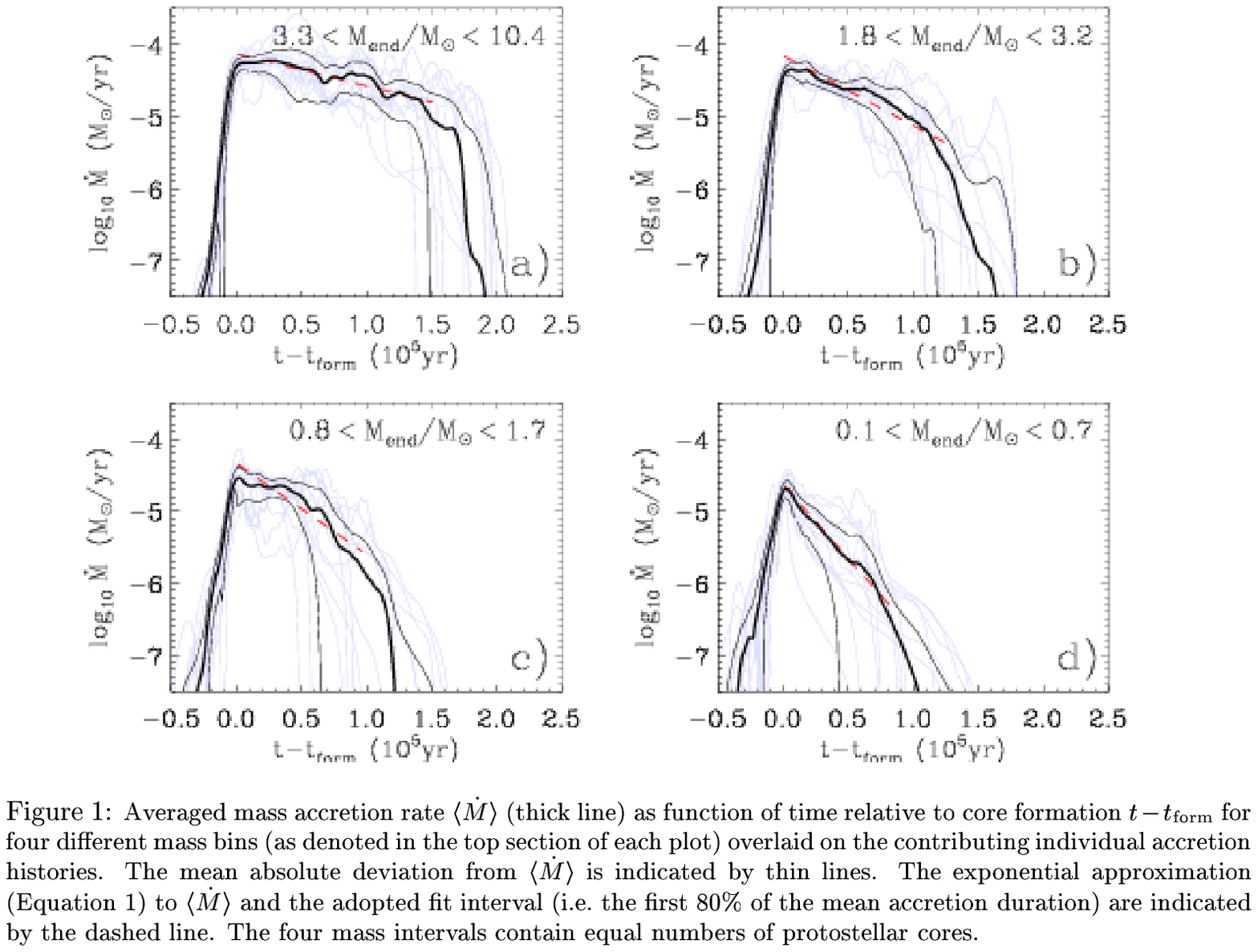}}
\end{picture}
\end{figure*}
\section{Time-Varying Protostellar Accretion Rates}
The following conclusions about the mass accretion rate in
dense clusters can be derived:

{\bf(1)} {\em Protostellar accretion rates in a dense cluster environment
  are strongly time variable}. This is  illustrated in Figure 1 for 49
randomly selected cores. 

{\bf(2)} The typical density profiles of gas clumps that give birth to
protostars exhibit a flat inner core, followed by a density fall-off
$\rho \propto r^{-2}$, and are truncated at some finite radius (in the
dense centers of clusters often due to tidal interaction with
neighboring cores). Figure 12 in KB1 presents examples.  As result, a
short-lived initial phase of strong accretion occurs when the flat
inner part of the pre-stellar clump collapses.  This corresponds to
the class 0 phase of protostellar evolution. If these cores were to
remain isolated and unperturbed, the mass growth rate would gradually
decline in time as the outer envelope accretes onto the center. This
is the class I phase.  Once the truncation radius is reached,
accretion fades and the object enters the  class II phase. This behavior is
expected from analytical models (e.g.\ HAB) and
agrees with other numerical studies (e.g.\ Foster \& Chevalier 1993,
WT).  However, collapse does not start from rest for the density
fluctuations considered here, and the accretion rates exceed the
theoretically predicted values even for the most isolated objects in
the simulation.

{\bf (3)} {\em The mass accretion rates of cores in a dense cluster
  deviate strongly from the rates of isolated cores. This is a direct
  result of the mutual dynamical interaction and competition between
  protostellar cores.} While gas clumps collapse to build up
  protostars, they may merge as they follow the flow pattern towards
  the cluster potential minimum. The timescales for both processes are
  comparable. The density and velocity structure of merged gas clumps
  generally differs significantly from their progenitor clumps, and
  the predictions for isolated cores are no longer valid.  More
  importantly, these new larger clumps contain multiple protostellar
  cores, which subsequently compete with each other for the accretion
  from a common gas reservoir. The most massive core in a clump is
  hereby able to accrete more matter than its competitors (Bonnell et
  al.\ 1997, KB1). Its accretion rate is {\em enhanced} through the
  clump merger, whereas the accretion rate of low-mass cores typically
  {\em decreases}.  Temporary accretion peaks in the wake of clump
  mergers are visible in abundance in Figure 1.  Furthermore, the
  small aggregates of cores that build up are dynamically unstable and
  low-mass cores may be ejected. As they leave the high-density
  environment, accretion terminates and their final mass is reached.
  
{\bf(4)} {\em The most massive protostars begin to form first and
  continue to accrete at high rate throughout the entire cluster
  evolution.} As the most massive gas clumps tend to have the largest
  density contrast, they are the first to collapse and constitute the
  center of the nascent cluster.  These protostellar cores are fed at high rate
  and gain mass very quickly. As their parental clumps merge with
  others, more gas is fed into their `sphere of influence'. They
  are able to maintain or even increase the accretion rate when
  competing with lower-mass cores (e.g.\ core 1 and 8 in Figure
  1).  Low-mass stars, on average, tend to form somewhat later in
  the dynamical evolution of the system (as indicated by the absolute
  formation times in Figure 1; also Figure 8 in KB1), and typically
  have only short periods of high accretion.

{\bf(5)} {\em Individual cores in a cluster environment form and
  evolve through a sequence of highly probabilistic events, therefore,
  their accretion histories differ even if they accumulate the same
  final mass.} Accretion rates for protostars of certain mass can only
  be determined in a statistical sense. For the investigated cluster,
  I define four mass bins each containing 14 cores, and calculate the
  average accretion rate $\langle {\dot{M}} \rangle$ and its mean
  absolute deviation.  The result is shown in Figure 2, with the
  considered mass range indicated at the top of each plot.  
  An exponential decline with a cut-off at $t_{\rm end}$ offers a
  reasonable approximation to $\langle {\dot{M}} \rangle$,
\begin{equation}
\log_{10}{\langle\dot{M}\rangle} \,\,\left[{\rm M}_{\odot}{\rm
yr}^{-1}\right] = A -\frac{t-t_{\rm form}}{\tau} \,\,\,\,\,\,\,{\rm{for}}\,\,\,\,\,\,
0 \le t-t_{\rm form} < t_{\rm end}\,\nonumber
\end{equation}
with normalization $A$ and decline time $\tau$  obtained from fitting $\langle
\dot{M} \rangle$ over 80\% of the average accretion duration.
There is an implicit upper mass limit $M_{\infty}= 10^A
\tau/\ln{\!10}\,\,\,$M$_{\odot}$ when $t_{\rm end}\rightarrow
\infty$. The integration of $\langle \dot{M} \rangle$  yields $\langle
M_{\rm end}\rangle < M_{\infty}$.  The values $A$, $\tau$, $\langle
M_{\rm end}\rangle$, and $M_{\infty}$ are
listed in Table 1. Note, as individual accretion rates deviate
considerably from the cluster average, when {\em constructing} cluster
accretion rates from the fit formula a time-dependent random component
with roughly 50\% deviation should be superimposed onto the
mean.


\begin{center}\begin{minipage}[h]{2cm}{\sc Table 1}\end{minipage}
\renewcommand{\arraystretch}{1.2}
\begin{tabular*}{8.4cm}[t]{@{\extracolsep{\fill}}ccccc}
\tableline
\tableline\\[-0.4cm]
{A (M$_{\odot}$yr$^{-1}$)} &
{$\tau$ ($10^5\,$yr)}  &
{$\langle M_{\rm end} \rangle$  (M$_{\odot}$)} &
{$M_{\infty}$ (M$_{\odot}$)} &
{bin}\\[0.1cm]
\tableline\\[-0.3cm]
${-4.13}$ & $2.24$ & $4.17$ & $7.21$   & (a)\\
${-4.16}$ & $1.03$ & $2.04$ & $3.09$   & (b)\\
${-4.35}$ & $0.79$ & $1.08$ & $1.54$   & (c)\\
${-4.66}$ & $0.49$ & $0.39$ & $0.47$   & (d)\\[0.1cm]
\tableline
\end{tabular*}
\begin{minipage}[t]{8.4cm}{Fit parameters to ${\langle \dot{M}
\rangle}$ applying Equation 1 to the initial 80\% of the mean
accretion duration in  mass bins (a) to (d). }\end{minipage}\end{center}

{\bf(6)} As $\dot{M}$ and $M_{\rm end}$ of evolving protostars in
dense clusters are influenced by mutual stochastic interactions, the
bolometric luminosity $L_{\rm bol}$ and temperature $T_{\rm bol}$ are
not functions of mass and age alone (Myers et al. 1998, WT), but also
depend on the statistical properties of the parental cluster (WK).
{\em For protostars in the main accretion phase, mass and age
determination from comparison with theoretical tracks, either in the
$\log T_{\rm bol}-\log L_{\rm bol}$ or the Hertzsprung-Russell (H-R)
diagram, is only possible in an approximated way -- as the average
over many different theoretical accretion histories for different
cluster environments.}  This may still affect protostars when they
enter the pre-main sequence contraction phase and accretion fades off.
A quantitative analysis of the expected errors for different cluster
environments will be the subject of a subsequent investigation.

\acknowledgements{I thank Peter Bodenheimer and G{\"u}nther Wuchterl
for many stimulating discussions and for their advice, and an
anonymous referee for detailed and insightful comments. }

\end{document}